\documentclass[11pt]{article}
\usepackage[margin=2cm]{geometry}

\usepackage{listings}
\usepackage{jheppub}

\usepackage{amsmath, amsfonts, amssymb}
\usepackage{comment}
\usepackage{graphicx}
\usepackage{psfrag}
\usepackage[usenames,dvipsnames,svgnames,table]{xcolor}
\usepackage{enumerate}
\usepackage{soul}
\usepackage{subfig}
\usepackage{dsfont}
\usepackage{verbatim}
\usepackage{tikz-cd}
 \usepackage{cleveref}
\usepackage{slashed}
\usepackage{bm}
\usepackage{physics}
\usepackage{subfig}

\usetikzlibrary{decorations.markings,decorations.pathmorphing}
\tikzset{->-/.style={decoration={markings,mark=at position #1 with {\arrow{>}}},postaction={decorate}}}
\tikzset{>=latex}

  \usepackage{color}
  \definecolor{dark-gray}{gray}{0.20}
  \definecolor{gray}{gray}{0.30}
  \definecolor{light-gray}{gray}{0.80}
  \definecolor{dark-red}{rgb}{0.7,0,0}
  \definecolor{dark-green}{rgb}{0.1,0.4,0}
  \definecolor{dark-blue}{rgb}{0.3,0.3,0.7}
  \definecolor{light-blue}{rgb}{0.8,0.8,1}

\usepackage{hyperref}
\usepackage{setspace}
\hypersetup{
	colorlinks=true,
	linkcolor=dark-blue,
	citecolor=dark-red,
	urlcolor=dark-green,
	linktoc=page
}
\newcommand{\be}{\begin{equation}}
\newcommand{\ee}{\end{equation}}

\usepackage{pdfpages}

\def\be{\begin{equation}}
\def\ee{\end{equation}}
\def\bea{\begin{eqnarray}}
\def\eea{\end{eqnarray}}

\renewcommand{\Im}{\text{Im}\,}
\renewcommand{\Re}{\text{Re}\,}

\DeclareMathOperator{\arctanh}{arctanh}

\def\simleq{\; \raise0.3ex\hbox{$<$\kern-0.75em
      \raise-1.1ex\hbox{$\sim$}}\; }
   \def\simgeq{\; \raise0.3ex\hbox{$>$\kern-0.75em
      \raise-1.1ex\hbox{$\sim$}}\; }

\numberwithin{equation}{section}

\title{
Holographic (Eternal) Inflation
}

\author{Thomas Hertog$^{1,2,3}$,}
\author{Jef Pauwels$^{4,5}$,}
\author{and Victoria Venken$^6$}
\affiliation{$^1$ Institute for Theoretical Physics, KU Leuven,
Celestijnenlaan 200D, B-3001 Leuven, Belgium}
\affiliation{$^2$Leuven Gravity Institute, KU Leuven, Celestijnenlaan 200D, B-3001 Leuven, Belgium}
\affiliation{$^3$DAMTP, University of Cambridge, Wilberforce Road, Cambridge CB3 0WA, United Kingdom}
\affiliation{$^4$ Department of Applied Physics, University of Geneva, \\ Rue de l'Ecole-De-Médecine 20
1205 Genève, Switzerland}
\affiliation{$^5$ Constructor Institute of Technology, Rheinweg 9 8200 Schaffhausen, Switzerland}
\affiliation{$^6$ Rudolf Peierls Centre for Theoretical Physics, University of Oxford\\ Beecroft Building, Clarendon Laboratory, Parks Road, OX1 3PU, UK}
\emailAdd{thomas.hertog@kuleuven.be}
\emailAdd{jef.pauwels@unige.ch}
\emailAdd{victoria.venken@physics.ox.ac.uk}

\abstract{ 
We identify a minisuperspace of complex deformations of ABJM theory for which the partition function specifies the amplitude of an eternally inflating universe. The boundary theory predicts that the bosonic bulk is effectively in the Hartle-Hawking no-boundary state, with a subleading 'tunneling' contribution. This holographic model of inflation also reveals a close connection between the swampland distance and cobordism conjectures, and the condition for the asymptotic wave function to predict classical behavior in geometry and fields.  
}

\begin{document}

\maketitle

\newpage
\section{Introduction}

Eternal inflation refers to a near de Sitter regime deep in inflation, in which the quantum fluctuations in the energy density of the inflaton(s) are large. In the usual account of eternal inflation the quantum diffusion dynamics of fluctuations is modeled as a stochastic process around a classical slow-roll background. Since the stochastic effects dominate over the classical evolution it is argued eternal inflation produces a universe that is highly irregular on the largest scales, with exceedingly large or even infinite constant density surfaces \cite{Linde:1996hg}. 

However, this account is questionable, because the dynamics of eternal inflation wipes out the separation into classical backgrounds and quantum fluctuations that is assumed. A proper treatment of eternal inflation must be based on quantum cosmology. As a step towards this, we put forward a holographic model of scalar field driven eternal inflation using gauge-gravity duality.

\begin{figure}[t]
\begin{center}
\includegraphics[width=4in]{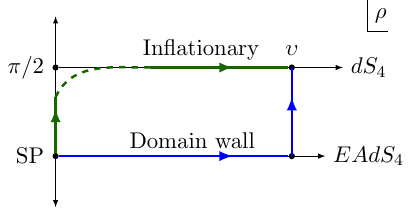} 
\caption{Two representations in the complex time-plane of the same no-boundary saddle point associated with an inflationary universe. The saddle point action includes an integral over time $\rho$ from the no-boundary origin or South Pole (SP) to its endpoint $\upsilon$ on a spacelike surface $\Sigma_f$. Different contours for this give different geometric representations of the saddle point, each giving the same amplitude for the final real configuration $(h_{ij}(\vec x),\phi (\vec x))$ on $\Sigma_f$. The saddle point geometry along the contour going horizontally from the SP consists of a regular, Euclidean, locally AdS domain wall with a complex scalar profile. Its regularized action specifies the tree-level probability in the no-boundary state of the associated inflationary, asymptotically de Sitter history that emerges on the upper nearly horizontal contour. Euclidean AdS/CFT relates the tree-level weighting to the partition function of a dual field theory, yielding \eqref{dSCFT}. }  
\label{contour}
\end{center}
\end{figure}

Gauge-gravity duality, or dS/CFT in this cosmological context \cite{Balasubramanian2001,Strominger2001}, conjectures that the no-boundary wave function of the universe on a spacelike surface $\Sigma_f$ is given in the large three-volume regime in terms of the partition function of certain deformations of a Euclidean CFT living on the future boundary \cite{Maldacena2002,Harlow2011,Hertog2011,Maldacena:2011nz,Anninos2011}. Further, Euclidean AdS/CFT generalized to complex relevant deformations implies an approximate, semiclassical realization of dS/CFT. This follows from the observation \cite{Hertog2011} that {\it all} no-boundary saddle points in low energy gravity theories with a positive scalar potential $V$ admit a geometric representation in which their weighting is fully specified by the regularized action of an interior, locally AdS, domain wall region governed by an effective negative scalar potential $-V$. We illustrate this correspondence in Fig. \ref{contour}. Schematically, gauge-gravity duality in a cosmological setting reads \cite{Hertog2011}, 
\be 
\Psi [h_{ij}, \phi]= Z^{-1}_{QFT}[\tilde h_{ij},\zeta] \exp(iS_{st}[h_{ij}, \phi]/\hbar)   \ .
\label{dSCFT}
\ee
Here the external sources $(\tilde h_{ij}, \zeta)$ in the dual, where $\zeta$ can be complex, are locally related to the argument $(h_{ij}, \phi)$ of the wave function $\Psi$. The phase factor is specified by the usual surface terms $S_{st}$, and the QFTs in this form of dS/CFT are complex deformations of Euclidean AdS/CFT duals. The boundary metric $\tilde h_{ij}$ stands for background {\it and} fluctuations. The reason the inverse of the AdS/CFT dual partition function enters in \eqref{dSCFT} can be traced to the fact that the Hartle-Hawking wave function in cosmology is related to the decaying wave function in AdS, whereas standard Euclidean AdS/CFT is concerned with the growing branch of the AdS wave function \cite{Gabriele:2015gca}.

This formulation of the duality lends some support to the view that Euclidean AdS/CFT and dS/CFT aren't so much analytic continuations of each other but rather two real sections of a single complexified theory \cite{Maldacena2002,Hull:1998vg,Dijkgraaf:2016lym,Bergshoeff:2007cg,Skenderis:2007sm,Hartle2012b}. The recent result that the leading and subleading quantum corrections to the dS entropy in four dimensions can be computed independently on the sphere, on EAdS and in the ABJM dual suggests that this view may hold true beyond the semiclassical approximation \cite{Bobev:2022lcc}. 

The proposed form \eqref{dSCFT} of dS/CFT and variations thereof have led to a fruitful application of holographic techniques to early universe and quantum cosmology (see e.g. \cite{Strominger:2001gp,McFadden2009,Bzowski:2012ih,Maldacena:2011nz,Garriga:2014fda,Afshordi:2016dvb}). Even though no dual field theories have been identified that correspond to top-down models of realistic cosmologies, where inflation transitions to a decelerating phase, we find that some of the known AdS/CFT duals are ideally suited to study eternal inflation from a holographic viewpoint. This is because supergravity theories in $AdS_4$ typically contain scalars with negative mass $m^2=-2l^2_{AdS}$ around a negative maximum, and with a negative potential for large $\phi$. In the context of \eqref{dSCFT} such scalars give rise to (slow roll) eternal inflation in the dS domain of the theory that is governed effectively by $-V$. In fact the Breitenlohner-Freedman bound in AdS corresponds precisely to the condition for eternal inflation in de Sitter.

As a specific example of a top-down model this kind, we consider homogeneous mass deformations of $k=1$ ABJM theory \cite{Aharony:2008ug} on $S^3$ that preserve $\mathcal{N}=1$. The partition function for real values of the sources turning on the deformations was calculated in \cite{Benna:2008zy,Kapustin:2009kz,Drukker:2010nc,Jafferis:2011zi}. It was matched to leading order to the action of bulk saddle points in a consistent truncation of Euclidean supergravity in four dimensions whose bosonic sector involves gravity coupled to six independent complex scalars \cite{Freedman:2013oja}. Here we consider this dual pair for complex values of the sources. Specifically, we identify a one-parameter set of complex mass deformations of ABJM for which the interior of the corresponding bulk saddle point describes the birth of an inflationary universe in the no-boundary state. The single free parameter is related to the value of the source in the dual and governs the amount of scalar field inflation in the bulk, before the inflaton rolls down to the minimum of the scalar potential $-V$. The dependence of the partition function on the value of the external source specifies a holographic no-boundary measure on this minisuperspace of homogeneous and isotropic, inflationary, asymptotically de Sitter universes. 
In this minisuperspace the six independent complex scalars in the bulk theory collectively act as an effective single inflaton rolling down in a positive effective $\cosh$ potential. Nevertheless, we find that the holographic no-boundary measure in this model differs slightly from the standard no-boundary measure in a $\cosh$ potential. We trace this difference to an additional surface term in the theory that is required by supersymmetry. 

A key feature of the standard no-boundary wave function in a $\cosh$-model is that it severely constrains the configuration space of inflaton values \cite{Hartle:2008ng,Dorronsoro_2017}. Specifically, the semiclassical wave function vanishes for field values at the South Pole of the saddle that are larger than a certain threshold value where the potential becomes steep and the conditions for inflation no longer hold. To be precise, while there exist no-boundary saddles for larger field values, they do not describe the birth of an asymptotically classical universe, and consequently their action diverges \cite{Hartle:2008ng,Dorronsoro_2017}. It is therefore an interesting question whether the configuration space of deformations that specifies the holographic wave function exhibits similar constraints. We find that not only it does, but that the configuration space implied by the holographic dual coincides precisely with the region of minisuperspace where the usual no-boundary wave function predicts classical behavior of geometry and fields. Furthermore, the range of field values that this selects in this model is roughly the Planck scale, so these restrictions resonate with swampland considerations \cite{Ooguri:2006in,Klaewer:2016kiy,Agrawal:2018own}. We thus arrive at an intriguing connection between the swampland distance conjecture, and the criterion for the wave function of the universe to predict classical behavior in geometry and fields in the first place.  

The most surprising property of the holographic wave function in this model is perhaps that it predicts, at least with the fermionic sector integrated out, a subleading contribution to the bulk wave function with the characteristics of the tunneling proposal. This surprising feature can be traced to the fact that the dual adopts so-called mixed boundary conditions. The net result is that the usual no-boundary saddle provides the dominant contribution to the wave function, albeit with a slightly altered weighting due to the SUSY surface term, but that there is a subleading 'tunneling' contribution. 

A reliable theory of eternal inflation is important to sharpen the predictions of slow roll inflation. This is because the physics of eternal inflation specifies a prior over the so-called zero modes, or classical slow roll backgrounds, which in turn determines the predictions for the precise spectral properties of CMB fluctuations on observable scales. So, despite the term eternal inflation and the fact that the dual evaluates the wave function in the large volume regime, one should really think of the holographic model we put forward as a theory of initial conditions. Loosely speaking we envisage employing holography to excise the bulk regime of eternal inflation and replace this by field theory degrees of freedom on a kind of end-of-the-world brane that is located, say, at the transition to slow roll inflation. This is somewhat analogous to the holography iof vacuum decay in AdS \cite{Maldacena:2010un}.

\section{Inflationary Saddles from Complex EAdS Domain Walls}
\label{sectionholographicinflation}
In this section we identify a set of complex generalizations of the BPS solutions of Euclidean 4D supergravity found by Freedman and Pufu (FP) \cite{Freedman:2013oja} that describe the birth of an inflationary universe in the no-boundary state.  

 FP considered the consistent truncation of the supergravity dual to $k=1$ ABJM theory \cite{Aharony:2008ug}. The bosonic part of the $\mathcal{N}=1$ Euclidean 4D supergravity action is given by\footnote{We set $8 \pi G_N = 1$.}
\begin{equation} \label{eq:FPaction}
	S_\text{bulk} = \int \dd^4x\, \sqrt{g} \left[ - \frac 12 R + \sum_{\alpha = 1}^3 \frac{\partial_\mu z^\alpha \partial^\mu \tilde z^\alpha }
    {\left(1 - z^\alpha \tilde z^\alpha \right)^2} 
   +\frac{1}{L^2} \left( 3 - \sum_{\alpha = 1}^3 \frac{2}{1 - z^\alpha \tilde z^\alpha} \right)  \right]\,,
\end{equation}
where $g$ is the determinant of the metric, $R$ is the Ricci scalar, $L = 3/ |\Lambda|$ is the AdS length and $z^\alpha, \tilde{z}^\alpha$ are \emph{six} independent \emph{complex} scalar fields. In a Lorentzian AdS setting $z^\alpha$ and $\tilde{z}^\alpha$ are required to be conjugate, but there is no such requirement in the Euclidean.
FP found two branches of everywhere real BPS solutions to \cref{eq:FPaction}, given by
\begin{align} \label{eq:FPsolutionsbranch1}
    ds^2 &= \frac{4 L^2 (1 + c_1 c_2 c_3) (1 + c_1 c_2 c_3 r^4) }{(1-r^2)^2 (1 + c_1 c_2 c_3 r^2)^2} \left( \dd r^2 + r^2 \dd \Omega_3^2 \right) \,, \\
    z^\alpha &= \frac{c_\alpha (1 - r^2)}{1 + c_1 c_2 c_3 r^2} \,, \qquad
      \label{eq:FPsolutionsbranchscalar}\tilde z^\alpha = - \frac{c_1 c_2 c_3 (1 - r^2)}{c_\alpha ( 1 + c_1 c_2 c_3 r^2) } \,.
\end{align}
and a second branch with $z^\alpha$ and $\tilde{z}^\alpha$ interchanged. In the above solutions, the three constants $c_\alpha$ are real and the conformal AdS boundary is located at $r=1$. In terms of a new coordinate $\rho \equiv 2 \arctanh(r)$, the metric \cref{eq:FPsolutionsbranch1} reads
\begin{equation}
    ds^2 = \frac{4 L^2 (1 + c_1 c_2 c_3) (1 + c_1 c_2 c_3 \tanh^4 (\rho/2)) }{(1-\tanh^2 (\rho/2))^2 (1 + c_1 c_2 c_3 \tanh^2 (\rho/2))^2} \left( \frac{1}{4} \text{sech}^4\left(\frac{\rho}{2}\right) \dd \rho^2 + \tanh^2 (\rho/2) \dd \Omega_3^2 \right) \,.
\label{eq:metricrhofull}
\end{equation}
Near the boundary one has
\begin{equation} \label{eq:FPnearboundarycoordinate}
	 r = 1 - 2 e^{- \rho} + 2 e^{-2\rho} - 2 \frac{(1 - c_1 c_2 c_3)^2}{(1 + c_1 c_2 c_3)^2} e^{-3 \rho} + \ldots \,,
\end{equation}
Hence the asymptotic metric and fields are given by
\begin{equation} \label{eq:FPnearboundarymetric}
  ds^2 = L^2 \dd \rho^2 + \frac {L^2 e^{2 \rho}}4 \left(1 - \frac{1 + c_1 c_2 c_3 (c_1 c_2 c_3 - 10)}{(1 + c_1 c_2 c_3)^2} e^{-2 \rho} + \cdots \right)^2 \dd \Omega_3^2 \,,
\end{equation}
and
\begin{align}
\label{eq:FPnearboundaryscalar1}
  z^\alpha(\rho) &= \frac{4 c_\alpha}{1 + c_1 c_2 c_3} e^{-\rho} - \frac{8 c_\alpha (1- c_1 c_2 c_3)}{(1 + c_1 c_2 c_3)^2} e^{-2 \rho} + \cdots \,, \\ \label{eq:FPnearboundaryscalar2}
  \tilde z^\alpha(\rho) &=  -\frac{4 c_1 c_2 c_3}{c_\alpha (1 + c_1 c_2 c_3)} e^{-\rho} 
     + \frac{8 c_1 c_2 c_3 (1- c_1 c_2 c_3)}{c_\alpha(1 + c_1 c_2 c_3)^2} e^{-2 \rho}+ \cdots \,.
\end{align}
The FP solutions describe regular, real domain walls that are asymptotically Euclidean AdS.   
The field theory dual to these saddle points are real mass deformations of ABJM theory on $S^3$ \cite{Freedman:2013oja}. 

We are interested in complex extensions of the FP solutions. These can be viewed as solutions in the complex $\rho$-plane. Taking $\rho$ complex as in \cref{contour} we see there is a curve $\rho = i\pi/2 + t$, with $t$ real, along which, for increasing $t$ the asymptotic metric tends to leading order to Lorentzian de Sitter space.
Indeed, substituting $\rho = i\pi/2 + t$ in \cref{eq:FPnearboundarymetric} yields
 \begin{equation} 
\label{eq:FPnearboundarymetricdS}
  ds^2 = L^2 \dd t^2 - \frac {L^2 e^{2 t}}4 \left(1 + \frac{1 + c_1 c_2 c_3 (c_1 c_2 c_3 - 10)}{(1 + c_1 c_2 c_3)^2} e^{-2 t} + \cdots \right)^2 \dd \Omega_3^2 \,,
\end{equation}
However, the asymptotic scalar fields in the FP solutions above are to leading order imaginary for large $t$. Hence we must generalize the FP solutions to complex scalar profiles along the EAdS domain wall, for real values of $\rho$, in order to obtain real inflationary cosmologies along the de Sitter curve $\rho = i\pi/2 + t$ in the complex $\rho$-plane. 

\begin{figure}[h!]
	\centering
	\includegraphics[width=0.8\textwidth]{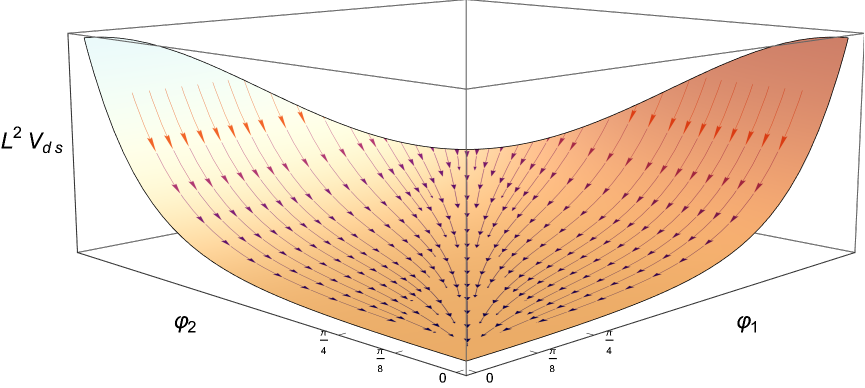}
	\caption{Two-dimensional slice of the scalar potential, $\frac{L^2}{3}V(\varphi_1,\varphi_2) = \cosh(\sqrt{\frac{2}{3}}\varphi_1) + \cosh(\sqrt{\frac{2}{3}}\varphi_2)$. Superimposed in a flow field for the gradient. The BPS-line where $\varphi_1=\varphi_2$ is an attractor. The $V_{dS}$-axis is in units of $L^2$.\label{fig:2Dinflationpot}}
\end{figure}

Anticipating the identification of a combination of the scalars $z^\alpha, \tilde{z}^\alpha$ with an inflaton, let us reparametrize the fields as follows,
\begin{align}
	z^\alpha &= \tanh( \varphi_\alpha /\sqrt{6})e^{i\theta_\alpha}, \\
	\tilde{z}^\alpha &= \tanh(  \varphi_\alpha/\sqrt{6})e^{-i\theta_\alpha}.
\end{align}
where the six fields $\theta_\alpha,\varphi_\alpha$ take values in $\mathds{C}$. 
In terms of $\theta_\alpha,\varphi_\alpha$ the matter Lagrangian is given by
\begin{equation} \label{eq:FPeffectivematter}
	\mathcal{L}_{\rm mat}  =  \sqrt{g}\sum_{\alpha=1}^3 \qty[ \frac{1}{6}\partial_\mu \varphi_\alpha \partial^\mu \varphi_\alpha + \cosh^4(\varphi_\alpha/\sqrt{6}) \tanh^2 (\varphi_\alpha/\sqrt{6}) \partial_\mu \theta_\alpha \partial^\mu \theta_\alpha + V(\varphi_\alpha)] ~,
\end{equation}
where
\begin{equation} \label{eq:FPinflatonpotential}
	 V(\varphi_\alpha) = -\frac{1}{L^2}  \cosh(\sqrt{\frac{2}{3}} \varphi_\alpha) ~.
\end{equation}
Hence the three fields $\varphi_\alpha$ enter with a canonical kinetic term in the action, and with an identical $\cosh$ potential.
Also, the scalar potential $V$ is independent of the fields $\theta_\alpha$, which are merely constant spectator fields in the FP solutions that do not affect the metric. 
SUSY heavily restricts the solution space indeed. 
This is evident from the metric in \cref{eq:FPsolutionsbranch1}, which only depends on the product $\prod_\alpha c_\alpha$ and not on the individual constants $c_\alpha$ that specify each separate profile $\varphi_\alpha$. Likewise, the product $z^\alpha \tilde{z}^\alpha$
is independent of $\alpha$. We can therefore define an effective field $\varphi$, 
\begin{equation}
	z^\alpha\tilde{z}^\alpha = \tanh^2 (\varphi_\alpha /\sqrt{6}) \equiv \tanh^2  (\varphi/\sqrt{6}),
\end{equation}
that governs the entire dynamics of the FP solutions, and that will act as \emph{one} effective inflaton field in the complex generalizations that are of cosmological interest. 
As mentioned in the Introduction, along the asymptotic de Sitter curve in the complex $\rho$-plane, the scalars feel an effective potential $V_{dS} = -V$. In fact, looking at the full potential for the three scalars $\varphi_\alpha$ along the de Sitter branch, i.e. $ \frac{L^2}{3}V_{dS}(\varphi_1,\varphi_2,\varphi_3) = \sum_\alpha \cosh( \sqrt{\frac{2}{3}}\varphi_\alpha)$, we see that SUSY selects solutions with $\varphi_1 =\varphi_2 =\varphi_3$ that flow along the flattest potential direction, which is an attractor indeed. A two-dimensional slice through field space depicting this is given in \cref{fig:2Dinflationpot}.

In essence, therefore, the FP solutions and their complex generalizations can be regarded as extrema of a reduced action containing a single scalar with Lagrangian, 
\begin{equation} \label{eq:reducedmatterlagrangian}
	 	\mathcal{L}_{\rm mat}  = \sqrt{g}\qty( \frac{1}{2}\partial_\mu \varphi \partial^\mu \varphi   -\frac{3}{L^2}  \cosh( \sqrt{\frac{2}{3}} \varphi)) ~,
\end{equation}
To identify saddles that describe the birth of an inflationary universe, it remains to determine the phase of the complex parameters $c_\alpha$ for which $\varphi$ becomes approximately real along the Lorentzian dS contour in n \cref{contour}.
In terms of the original fields, we have
\begin{equation}\label{eq:physicalscalerdef}
		\varphi = \sqrt{6}\;\text{atanh}(\pm \sqrt{z^\alpha \tilde{z}^\alpha}) ~,
\end{equation}
Using \cref{eq:FPnearboundaryscalar1} and \cref{eq:FPnearboundaryscalar2}, we obtain the Fefferman-Graham expansion of $\varphi$ along the dS contour,
\begin{equation} \label{eq:inflatondS}
  \frac{\varphi(t)}{\sqrt{6}} = \pm \frac{4 \sqrt{\prod_\alpha c_\alpha}}{1+ \prod_\alpha c_\alpha} e^{-t} \mp \frac{8\prod_\alpha c_\alpha(1-\prod_\alpha c_\alpha)}{(1+\prod_\alpha c_\alpha)^2}e^{-2t} \pm \mathcal{O}(e^{-3t}) ~.
\end{equation}
Hence we see that $\varphi (t) $ is real to leading order $\sim e^{-t}$ provided the combination 
\begin{equation} \label{eq:constraintFP1}
	\varphi_B \equiv \pm \sqrt{6} \frac{4 \sqrt{c_1 c_2 c_3}}{(1+c_1c_2c_3)} ~,
\end{equation}
is real. This can be achieved either by taking $c \equiv c_1 c_2 c_3$ real and positive or by taking $c$ complex and $||c|| =1$. 
The former option covers the range $[-2 \sqrt{6}, 2\sqrt{6}]$ of values of the coefficient \cref{eq:constraintFP1}  
Note also that the coefficient of the leading term in the asymptotic expansion \cref{eq:inflatondS} is invariant under $c \rightarrow 1/c$. Hence the range of values of $c$ restricted to either $[0,1]$ or $[1,+\infty]$ covers the same asymptotic boundary profiles for $\varphi$, albeit with different subleading behaviour. Below we will see that also the saddle-point action is real for this branch of solutions and obeys $I \xrightarrow{c\rightarrow 1/c} -I$. The latter option of taking $c$ complex with $||c|| =1$ covers the remaining range of values of the coefficient \cref{eq:constraintFP1}, i.e. the union of $[-\infty,-2 \sqrt{6}]$ and $[2 \sqrt{6},+\infty]$. The invariance of the coefficient under $c \rightarrow 1/c$ means that the upper and lower half unit circle in the complex $c$-plane cover the same range of asymptotic boundary value of $\varphi$. Below we will see that the action of these saddles is purely imaginary and again obeys $I \xrightarrow{c\rightarrow 1/c} -I$.

\begin{figure}[t!]
    \centering
    \includegraphics[width=0.4\linewidth]{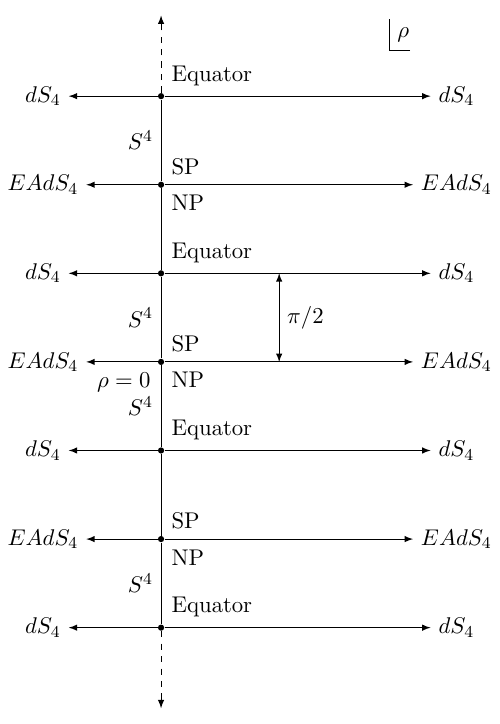}
    \caption{Shown is the complex geometry of the BPS solution with the scalars turned off. Along the imaginary $\rho$-axis the solution describes a tower of four spheres. Parallel to the real $\rho$-axis one has lines where the solution is Euclidean $AdS_4$ alternating with  lines along which it is Lorentzian $dS_4$, separated by a coordinate distance $\pi/2$ corresponding to half a four sphere.}
    \label{fig:emptygeo}
\end{figure}

Given that we have the complete solutions \cref{eq:metricrhofull}, we can easily analyze the geometry and fields in the entire complex $\rho$-plane. When the scalars are turned off, i.e. $c_1 c_2 c_3 =0$, the structure is rather trivial, featuring horizontal curves along which the geometry is pure Euclidean $AdS_4$, separated by half a four sphere moving along the imaginary $\rho$-axis, from a horizontal curve along which the geometry is Lorentzian de Sitter. This structure is shown in \cref{fig:emptygeo}.

\begin{figure}[t!]
\centering
\subfloat[$c_1 c_2 c_3 =1$]{\label{fig:c1_geo}
\centering
\includegraphics[width=0.30\linewidth]{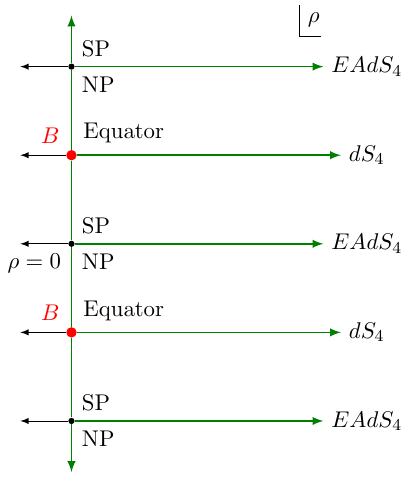}
}
\hfill
\subfloat[$c_1 c_2 c_3 <1$]{\label{fig:HH_geo}
\centering
\includegraphics[width=0.30\linewidth]{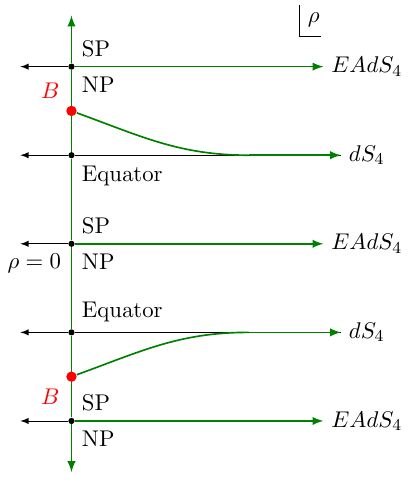}
}
\hfill
\subfloat[$c_1 c_2 c_3 >1$]{\label{fig:LV_geo}
\centering
\includegraphics[width=0.30\linewidth]{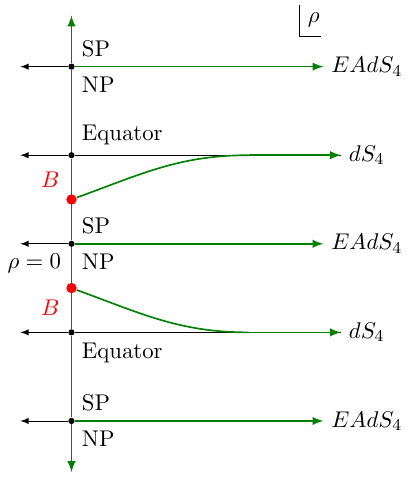}
}
\caption{Shown in green are the curves in the complex $\rho$-plane along which the metric component $g_{\rho \rho}$ is real, for real and positive values of $c\equiv c_1c_2c_3$. Points where $g_{\rho \rho}$ is singular are shown in red. The pattern repeats periodically for increasing $\Im{\rho}$ and the structure in the $\Re{\rho}<0$ half-plane mirrors that shown here.}
 \label{fig:realcgeo}
\end{figure}

Turning on the scalars brings about an important change in the complex geometry. Consider first the special case $c_1 c_2 c_3 =1$. 
Looking at the key metric component $g_{\rho \rho}$ in \cref{fig:realcgeo} (a), we find this is real precisely along the previous $EAdS_4$, $dS_4$, and $S^4$ contours. However, a new feature appears in that $g_{\rho \rho}$ diverges at the equator of the $S^4$. We label this point $B$. Eq. \cref{eq:FPsolutionsbranchscalar} shows that the scalars $z^\alpha$ and $\tilde{z}^\alpha$ also diverge at $B$, but the canonical scalar $\varphi$ as well as the Ricci curvature $R$ remain finite. Consequently, the saddle-point action remains finite and well-behaved. 
In fact, the canonical inflaton field $\varphi$ is purely imaginary on the $S^4$, so we can interpret this as a real scalar field $\bar{\varphi} \equiv  i \varphi$ moving in the effective potential 
\begin{equation}
V_{\rm S^4}(\bar{\varphi})= - \frac{3}{L^2} \cos(\sqrt{\frac{2}{3}}\bar{\varphi})	~.
\end{equation}
At $B$ one has $|\bar{\varphi}|=\sqrt{3/2}\pi$, which is precisely where the cosine potential reaches its maximum value.

Let us now track how the point $B$, where $|\bar{\varphi}|=\sqrt{3/2}\pi$ and $g_{\rho \rho}$ diverges, moves in the complex $\rho$-plane  when we vary $c_1 c_2 c_3$.

Consider first the case $c_1 c_2 c_3 < 1$. The corresponding behavior of $g_{\rho \rho}$ is shown in \mbox{\cref{fig:realcgeo} (b)}. One sees that the location of $B$ has moved away from the equator, towards the north (south) pole of the upper (lower) $S^4$. The point $B$ remains on the $S^4$, but $g_{\rho \rho}$ is no longer real along the entire horizontal line which asymptotes to $dS_4$. Instead, the curve along which $g_{\rho \rho}$ is real starts at $B$ and asymptotes to the horizontal $dS_4$ line. For $c_1 c_2 c_3 \rightarrow 0$, corresponding to $\varphi_{B} \rightarrow 0$, the upper (lower) $B$ tends to the north (south) pole.
 although exactly at $c_1 c_2 c_3 = 0$, the divergent behaviour of $g_{\rho \rho}$ and $z^\alpha$, $\tilde{z}^\alpha$ vanishes.

\begin{figure}[t!]
\centering
\subfloat[Lower complex]{\label{fig:lowerss_geo}
\centering
\includegraphics[width=0.47\linewidth]{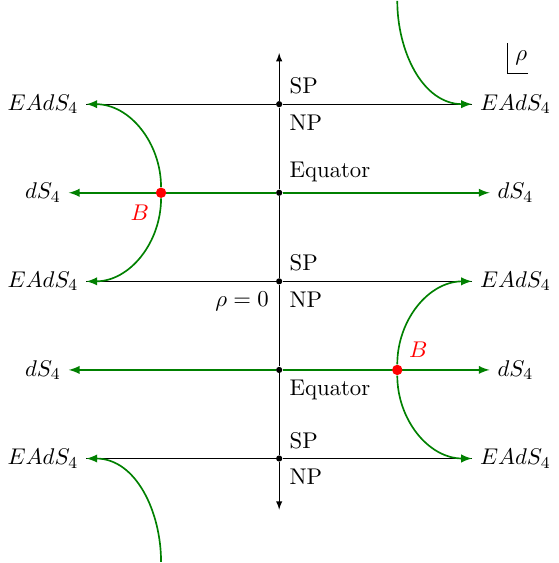}
}
\hfill
\subfloat[Upper complex]{\label{fig:uperss_geo}
\centering
\includegraphics[width=0.47\linewidth]{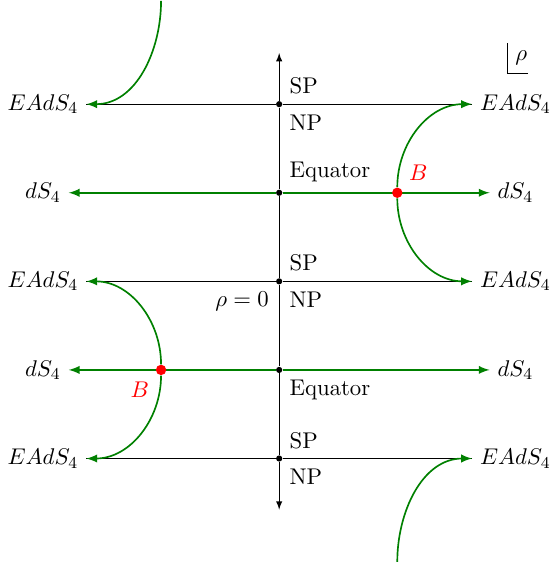}
}
\caption{Shown in green are the curves in the complex $\rho$-plane along which the metric component $g_{\rho \rho}$ is real, for two representative values of $c$ on the two hemispheres with $||c_1 c_2 c_3||=1$. Points where $g_{\rho \rho}$ is singular are shown in red. The pattern repeats periodically for increasing $\Im{\rho}$.}
\label{fig:phasecgeo}
\end{figure}

Finally we take $c_1 c_2 c_3$ complex with $||c_1 c_2 c_3|| = 1$, which is the remaining possibility to have a saddle exhibiting Lorentzian inflationary behavior. The slices along which the geometry is real in the complex $\rho$-plane are shown in \cref{fig:phasecgeo} (a) and \cref{fig:phasecgeo} (b) resp. for the upper (lower) semicircle. We see that $B$ is no longer shifted along the $S^4$ but along the $dS_4$ line, in opposite directions in the upper/lower half-plane. In fact, one no longer has a real $S^4$ geometry. Further, even the interior $EAdS_4$ geometry becomes complex. The line along which $g_{\rho \rho}$ is real passes through $B$ on the $dS_4$ line, curving towards the asymptotic $EAdS_4$ regime. In the limit $c_1 c_2 c_3 \rightarrow -1$, corresponding to $\varphi_{B} \rightarrow \pm \infty$, $B$ tends to $\pm \infty$.

In summary, the shape of the contours along which the geometry is real depends strongly on whether we consider the asymptotic Lorentzian class of solutions with $c_1 c_2 c_3$ real, or the complex $c$ solutions. In effect, this difference is a manifestation of far-reaching physical differences between the two classes of solutions. For $c_1 c_2 c_3$ real, and hence $|\varphi_B| \leq 2 \sqrt{6}$, the geometries are much like those of  familiar fuzzy instantons that describe the birth of an inflationary, asymptotically de Sitter universe in the no-boundary state. Specifically, there is a contour in the complex $\rho$-plane that connects the SP to the asymptotic de Sitter region, along which the geometry is everywhere regular and describes a transition from half a slightly deformed four sphere to a nearly Lorentzian, inflationary universe. 

By contrast, this is not the case for $||c_1 c_2 c_3||=1$, the class corresponding to $|\varphi_B| > 2 \sqrt{6}$, and several features indicate that these solutions are in fact not quite valid no-boundary saddles. First, this class of solutions does not exhibit a real $S^4$ regime. There isn't a real Euclidean on-shell geometry which dynamically shrinks the final boundary to nothing. Hence the asymptotic Lorentzian configuration isn't nucleated from nothing at the south pole, but rather from the finite boundary at $B$\footnote{The nucleation of an expanding universe from an initial boundary was recently studied in \cite{Friedrich:2024aad}. It would be interesting to investigate whether there is a connection between that proposal and the boundary $B$ featuring in our analysis.}. Put differently, the boundary is no longer dynamically cobordant to nothing, suggesting a contradiction with swampland principles, specifically the dynamical cobordism conjecture \cite{McNamara:2019rup,Buratti:2021yia,Buratti:2021fiv,Blumenhagen:2023abk}.

Second, also from the EAdS/CFT perspective these saddles are suspicious. Even though the geometries seem fine in the asymptotic Euclidean AdS regime, deep into the bulk the backreaction of the scalar yields a complex geometry which, at least for some range of $c$, violates the fluctuation convergence criterion recently revived by Kontsevich, Segal and Witten \cite{Kontsevich:2021dmb,Witten:2021nzp}. Specifically, while the KSW criterion is satisfied for all real $c$, it fails for complex $c$ with phase outside the interval $[-\pi/2,\pi/2]$. 

These considerations indicate that the class of complex $c$ saddles should be excluded and hence that the holographic wave function approximately 
vanishes in the large-field regime $|\varphi_B| > 2 \sqrt{6}$. In the next Section, we evaluate the action of both sets of solutions and further substantiate this.

\section{Holographic Wave Function}
The regularized action of the original asymptotically EAdS FP solutions \cref{eq:FPsolutionsbranch1} was calculated in \cite{Freedman:2013oja}. The regularization involves adding a number of counterterms $S_{ct}$ to the action \cref{eq:FPaction} that eliminate divergences in the large-volume limit in both the gravitational and scalar sectors. To preserve global supersymmetry, FP observed that the usual scalar counterterm must be slightly modified, with an additional asymptotically finite term that reads
\begin{equation}
\label{eq:SUSYboundaryterm}
S_{\rm SUSY} = \frac{1}{ L} \int_\partial d^3 x \sqrt{h}  \left[z_1 z_2 z_3 + \tilde{z}_1 \tilde{z}_2 \tilde{z}_3\right]\,.
\end{equation}

With this term added, the regularized on-shell Euclidean action $I_{reg}$ of the BPS solutions is given by
\begin{equation} \label{eq:FPactiononshell}
	I_{\rm reg }[c] = 4 \pi^2 L^2 \frac{1 - c}{1 + c} \,.
\end{equation}
Without the term $S_{\rm SUSY}$ added, the fraction in \cref{eq:FPactiononshell} would be raised to the third power and the action wouldn't reproduce the large $N$ dual field theory partition function. Once again we see that the on-shell action only depends on $c$, in line with our analysis of the geometry of the solutions. 

The action is single-valued as a function of $c$. However, to employ the holographic dictionary  \cref{dSCFT} we must compute the action as a function of the fields at $\upsilon$. We may rewrite the action as a function of $\varphi_B$.  We saw in \cref{sectionholographicinflation} that $c$ and $1/c$ provide two different solutions with the same value of $\varphi_B$. These two solutions have opposite actions. As a result, for given $\varphi_B$ one has two possible actions
\begin{equation} \label{eq:FPactiononshellphi}
	I_{\pm }[\varphi_B] = \pm 4 \pi^2 L^2 \sqrt{1- \left(\frac{\varphi_B}{2 \sqrt{6}}\right)^2} \,,
\end{equation}
where $I_+$ corresponds to $0 \leq c \leq 1$ for $0 \leq |\varphi|_B \leq 2 \sqrt{6}$ and $c$ on the lower unit complex half-circle for $|\varphi|_B > 2 \sqrt{6}$; while $I_-$ corresponds to $ c \geq 1$ for $0 \leq |\varphi|_B \leq 2 \sqrt{6}$ and $c$ on the upper unit complex half-circle for $|\varphi|_B > 2 \sqrt{6}$.

\begin{figure}[t!]
    \centering
    \includegraphics{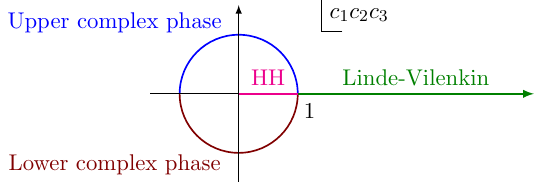}
    \caption{Shown are the values in the complex $c_1c_2c_3$-plane for which the FP saddles obey asymptotic de Sitter boundary conditions. The different colors discriminate between the behavior of the resulting semiclassical wave function, as discussed in the text.}     \label{fig:c1c2c3parameterspace}
\end{figure}

Now let's turn to the action of the complex generalizations of the FP solutions that are asymptotically de Sitter. To compute this one must integrate along either the green or blue contour in \cref{contour}  to the endpoint at $\upsilon$ where we evaluate the wave function. In this calculation one usually doesn't add counterterms. Yet the regularized action along the asymptotic EAdS leg of the contour does play an important role. What happens is that the action integral along the vertical part of the blue contour produces the usual gravitational and scalar counterterms $S_{ct}$, with a coefficient $(1+i)$, up to corrections that all decay in the large-volume limit \cite{Hertog:2011ky}. In particular it doesn't contribute to the asymptotically finite piece of the action. That said, given that we wish to maintain the connection with the dual, we include here too the finite surface term $S_{\rm SUSY}$ given by \cref{eq:SUSYboundaryterm}. We have verified that this term too is constant along the vertical contour leg for the complex FP solutions of interest. Hence the action integral along the vertical leg effectively performs holographic renormalization, canceling the large-volume divergences that arise from the horizontal, asymptotic EAdS leg of the contour, and producing a large phase factor in the wave function. The latter plays a role in determining whether the wave function predicts the configuration at $\upsilon$ will evolve classically. Further, this also means that, in the large-volume limit, the relative weighting of the different complex FP solutions that are asymptotically de Sitter is in fact given by the regularized action from the EAdS leg of the contour, with the $S_{\rm SUSY}$ contribution added. That is, the weighting is given by \cref{eq:FPactiononshell} where the coefficients $c_\alpha$ take the values identified in the previous Section for which the asymptotic dS regime emerges. The different curves in the complex $c$-plane selected by the asymptotic dS condition are summarized in \cref{fig:c1c2c3parameterspace}.

Substituting the action \cref{eq:FPactiononshellphi} in \cref{dSCFT} yields an expression for the semiclassical wave function $\Psi[\varphi_B]$ of the universe in this model. 
Interestingly, for a given value $\varphi_B$ of the effective scalar at the boundary surface $\upsilon$, which is conformal to the argument of the wave function, there are two saddle-point contributions. These two saddles correspond to inversely related $c$ values and hence differ in the subleading behavior of the effective scalar. They have opposite action $I_{\pm}$, as can be seen from \cref{eq:FPactiononshellphi}, resulting in a wave function of the form
\begin{equation}\label{wf}
    \Psi [\varphi_B] \sim  \left(\exp(I_{+ }[\varphi_B]) + \exp(I_{- }[\varphi_B])\right) \exp(i S_{st})\ .
\end{equation}

\begin{figure}[h!]
    \centering
    \includegraphics[scale=0.8]{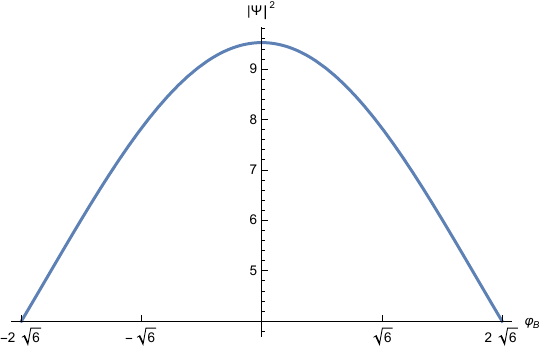}\includegraphics[scale=0.8]{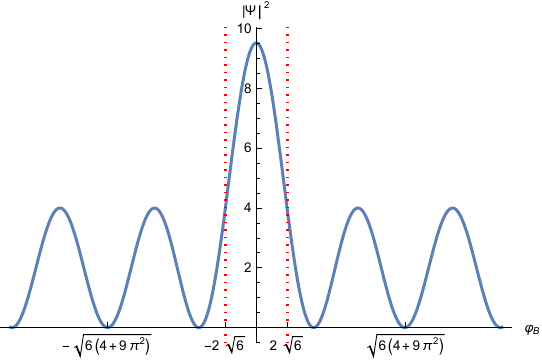}
    \caption{The value of $|\Psi|^2$ as a function of the inflaton value $\varphi_B$, with $L=1$. \textbf{Left:} The wave function squared exhibits the typical no-boundary behavior for small field values $|\varphi_B| < 2\sqrt{6}$. \textbf{Right:} The wave function behaves radically differently for large field ranges. The zeroes nearest to the origin are at $|\varphi_B| = 2 \sqrt{12}$, which is exactly the threshold beyond which the saddles don't satisfy the KSW criterion. For this and other reasons, we conjecture that the large-field saddles should be excluded and hence that the semiclassical wave function vanishes for large $\varphi_B$. The red dotted lines at $|\varphi_B| = 2\sqrt{6}$ indicate the limits of the small field ranges in panel (b). Note that the saddles contributing to $|\Psi|^2$ in the left panel enter with a weighting $\sim \exp(L^2 / l_p^2)$. Hence they are hugely enhanced, for $L \gg l_p$, relative to the saddles featuring in the right panel.}
    \label{fig:probdistroscalar}
\end{figure}

Given our discussion in Section \ref{sectionholographicinflation} it will not come as a surprise that the wave function behaves radically differently in the two distinct asymptotic dS regimes that we identified, resp. real $c \geq 0$ and complex $c$ with $||c|| =1$. For real $c \geq 0$, and hence $|\varphi_B| < 2\sqrt{6}$, the resulting wave function is a superposition of the familiar Hartle-Hawking and tunneling wave functions, with the latter providing merely an exponentially suppressed correction to the amplitudes\footnote{Wave functions of the universe that are a superposition of a Hartle-Hawking and tunneling contribution featured previously in \cite{Feldbrugge:2017mbc}, where it was argued that the tunneling contribution renders the wave function non-normalizable due to the enhancement of large perturbations. However, subsequently the tunneling contribution was shown to be absent after all in a number of exactly solvable models of this kind \cite{Janssen:2019sex,DiazDorronsoro:2018wro}. More recently, the combination of a Hartle-Hawking and tunneling instanton for lightly excited matter, and of an oscillatory wave function for highly excited matter, was also observed in a two-dimensional setting \cite{Anninos:2024iwf}.}. The shape of the probability distribution over $\varphi_B$ is illustrated in \cref{fig:probdistroscalar} (a) and exhibits the characteristic falloff for increasing $\varphi_B$.

For the range of complex $c$ with $||c|| =1$, corresponding to the large-field regime $|\varphi_B| > 2\sqrt{6}$, the action $I_{\rm reg}$ turns out to be purely imaginary. This yields a real oscillatory wave function in this regime, with a probability distribution $|\Psi|^2$ of the form illustrated in \cref{fig:probdistroscalar} (b). This behavior is very much unlike that of the Hartle-Hawking and tunneling wave functions, but it has been observed in higher-spin dS/CFT \cite{Anninos:2012ft,Hertog:2017ymy,Anninos:2017eib,Hertog:2019uhy}, notably in the case where the CFT has an equal number $N$ of fundamental and anti-fundamental scalars \cite{Hertog:2017ymy}. In this higher-spin dS/CFT context it was argued holographically that large source values should be excluded from the configuration space. Specifically, it was argued that the dual excludes boundary field values beyond the first zero of the wave function. 
In a somewhat similar vain, we have argued above that in the supergravity model we consider here, the solutions in the large-field regime aren't quite proper no-boundary saddles. The strange oscillatory behavior of the wave function lends further credence to the conclusion that the semiclassical wave function should indeed be taken to vanish for $|\varphi_B| \leq 2\sqrt{6}$. Although this sounds dramatic, note that since the amplitude of the Hartle-Hawking wave function is exponentially enhanced in the small-field regime, $\Psi_{HH} \sim \exp(L^2 / l_p^2)$, relative to the average amplitude in the oscillatory regime, the latter only slightly affects any predictions.

It is illuminating to consider the form of the small-field wave function \cref{wf} from a dual viewpoint. Remember that dS/CFT relates the argument of the wave function of the universe to sources in the dual. However in the model we consider the relation is more subtle. 
Schematically, the asymptotic Fefferman-Graham expansions, \cref{eq:FPnearboundaryscalar1} and \cref{eq:FPnearboundaryscalar2}, of the light scalars $z$ and $\tilde z$ are given by
\begin{align}
	 z^\alpha(\rho, x) &= a^\alpha(x) e^{- \rho} + b^\alpha(x) e^{-2 \rho} + \ldots \,, \\
  \tilde z^\alpha(\rho, x) &= \tilde a^\alpha(x) e^{- \rho} + \tilde b^\alpha(x) e^{-2 \rho} + \ldots \, ,
\end{align}
where $x$ is a coordinate on $S^3$. Usually $a^\alpha(x)$ and $\tilde{b}^\alpha(x)$ are canonically conjugate variables, and similarly for the tildes exchanged. Also, with normal boundary conditions in AdS/CFT, the leading coefficients in these expansions are identified with sources in the dual while the sub-leading coefficients determine the VEVs. However, there are two subtleties with this dictionary in this model, to do with the fact that $z$ and $\tilde z$ are light scalars and the presence of supersymmetry \cite{Freedman:2013oja}: Light scalars can have different boundary conditions, and supersymmetry relates boundary conditions of different scalars to each other. Specifically, the extra susy surface term implies that $a^\alpha$ is canonically conjugate to a quantity proportional to $\tilde{b}^\alpha - a^1 a^2 a^3 / a^\alpha$, and $\tilde{a}^\alpha$ is canonically conjugate to a quantity proportional to $b^\alpha - \tilde{a}^1 \tilde{a}^2 \tilde{a}^3 / \tilde{a}^\alpha$. The upshot is that in this model, dual sources are associated to
\begin{equation}
\label{eq:CFTsourcesgeneral}
a^\alpha - \tilde{a}^\alpha \qquad \text{and} \qquad \left(b^\alpha -\tilde{a}^1 \tilde{a}^2 \tilde{a}^3 / \tilde{a}^\alpha \right) + \left(\tilde{b}^\alpha - a^1 a^2 a^3 / a^\alpha \right)\,;
\end{equation}
and VEVs to
\begin{equation}
\label{eq:CFTVEVsgeneral}
a^\alpha + \tilde{a}^\alpha \qquad \text{and} \qquad \left(b^\alpha -\tilde{a}^1 \tilde{a}^2 \tilde{a}^3 / \tilde{a}^\alpha \right) - \left(\tilde{b}^\alpha - a^1 a^2 a^3 / a^\alpha \right)\,.
\end{equation}
Using \cref{eq:FPnearboundaryscalar1,eq:FPnearboundaryscalar2} one sees that these on-shell sources are invariant indeed under $c_\alpha \rightarrow 1/c_\alpha$, whereas the VEVs are not.  
Hence the superposition \cref{wf} of the Hartle-Hawking and tunneling wave functions is consistent from a dual viewpoint, in the sense that it corresponds to fixing the sources indeed. That said, one might expect that a full analysis of this model that includes e.g. the fermionic sector ultimately lifts this degeneracy. Even if that were the case, however, the above superposition would re-emerge when one considers probabilities for observables that coarse-grain over whatever source that discriminates between both saddles.

Finally, it is instructive to compare our results with the behavior of the no-boundary wave function in a 'bottom-up' minisuperpsace model consisting of Einstein gravity coupled to a positive cosmological constant and a single inflaton with a cosh potential.
The semiclassical Hartle-Hawking wave function in this setting was computed by Dorronsoro et al. in \cite{Dorronsoro_2017}, who also analyzed the conditions under which the wave function predicts late-time classical behavior. 
Working in a WKB spirit, these 'classicality conditions' boil down to the requirement that the amplitude of the wave function varies very slowly compared to its phase. When this is the case, the Wheeler-DeWitt equation predicts that the boundary configuration essentially evolves according to the equations of Lorentzian general relativity \cite{Hartle:2008ng}.

Interestingly, the region of minisuperspace for which \cite{Dorronsoro_2017} found the classicality conditions to hold, corresponds precisely to the range $|\varphi_B| \leq 2 \sqrt{6}$, in exact agreement\footnote{In \cite{Dorronsoro_2017}, the classicality conditions are evaluated at some $\upsilon$ in the large volume regime but not necessarily asymptotically, at future infinity. Their results agree with our findings in the late time limit, modulo a slight error. Specifically, \cite{Dorronsoro_2017} erroneously state that the condition for classicality, in their notation, is ${\phi}_1 < \sqrt{3/2 \lambda}$, where $\phi_1$ is the value of the inflaton at $\upsilon$ and $\lambda$ is a function of the scale factor at $\upsilon$, whereas the condition for classicality according to their calculations is actually ${\phi}_1 < \sqrt{3/ \lambda}$, in agreement with our results.} therefore with the bound we derived above on the basis of other considerations. 
However, whereas in bottom-up quantum cosmology, late-time classicality must be imposed by hand, holography in the top-down approach pursued here automatically implies a similar restriction on the configuration space of the wave function. Specifically, it turns out that the full range of deformations for which the dual partition function is reasonable and well-defined, covers the entire regime where the classicality conditions hold, but excludes the regime where the asymptotic wave function does not behave classically. Indeed for source values corresponding to $|\varphi_B| > 2 \sqrt{6}$, the partition function of the dual, Euclidean CFT is complex-valued rather than real. 
Even properties such as the R-charges of operators become complex in this regime. The R-charge is the charge of an operator under the $U(1)_R$ group. Since representations of this are only properly defined for real R-charge, this complex regime is group-theoretically ill-defined. From a bulk perspective, while the geometry in the gravitational dual is Euclidean AdS near the boundary $\rho \rightarrow + \infty$, the interior geometry is complex rather than Euclidean. For  $|\varphi_B| > 2 \sqrt{12}$ this complex behavior leads to a violation of the KSW criterion deep in the interior. Both sides of the duality, then, indicate that the saddles ought to be excluded.

In fact, this restriction on the inflaton range is entirely natural from an EAdS perspective. Remember that for real $c$, the effective inflaton $\varphi$ has an imaginary profile along the asymptotic EAdS leg of the contour in \cref{contour}, Hence it can be viewed as a real scalar field $\bar{\varphi} \equiv  i \varphi$ moving in the potential 
\begin{equation}
V_{\rm EAdS}(\bar{\varphi})= - \frac{3}{L^2} \cos(\sqrt{\frac{2}{3}}\bar{\varphi})	~.
\end{equation} 
The value of the inflaton field $\bar{\varphi}$ at the South Pole at $r=0$ is given by
\begin{equation} \label{eq:varrho0}
	\bar{\varphi_0} \equiv \bar{\varphi}(0) = \pm \sqrt{6}\;\text{atan}  \sqrt{c} ~.
\end{equation}
Hence this is constrained to the domain $|\varphi_0| \in [0,\sqrt{6}\pi/4]$ for the HH solution, and $|\varphi_0| \in [\sqrt{6}\pi / 4,\sqrt{6}\pi/ 2 ]$ for the tunneling solution\footnote{Note that this range of values at the South Pole corresponds precisely to the range $|\varphi_B| \leq 2 \sqrt{6}$ of conformal boundary values.}.

\begin{figure}[h!]
  \centering
  \includegraphics[width=0.8\textwidth]{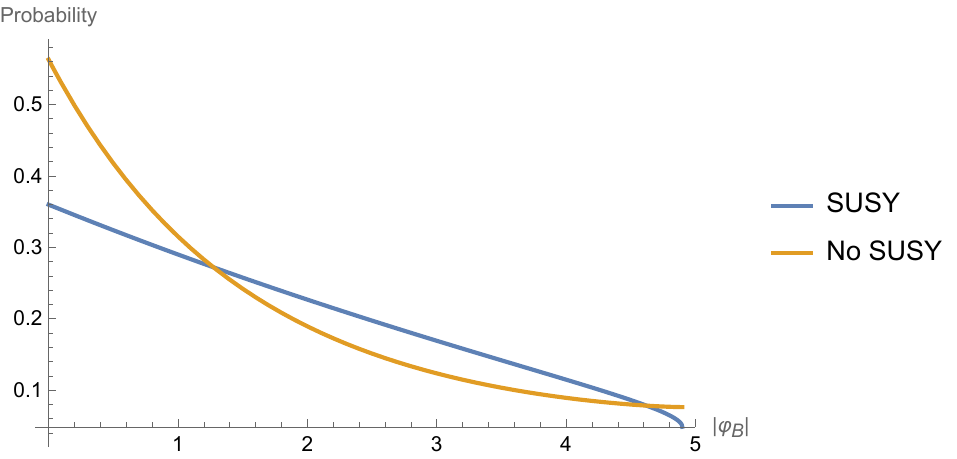}
  \caption{Comparison of the holographic no-boundary measure over inflationary universes, labeled by the boundary value $\varphi_B$, with the standard no-boundary measure in a $\cosh$-potential. Probabilities are normalised over the domain in which the no-boundary wave function predicts classical histories. \label{fig:NBmeasure}}
\end{figure}

Finally, we briefly discuss the holographic no boundary measure $\abs{\Psi_\text{HH} (\varphi_B)}^2$ in this model, on the sub-ensemble of asymptotically classical, inflationary universes. The Hartle-Hawking saddle has action $I_+$ in \cref{eq:FPactiononshellphi}. As discussed earlier, this receives a contribution from a finite supersymmetric surface term \cref{eq:SUSYboundaryterm}. Without the supersymmetric surface term, the action would be given by
\begin{equation} 
	I_{+\; \text{non-SUSY} }[\varphi_B] = 4 \pi^2 L^2 \left(1- \left(\frac{\varphi_B}{2 \sqrt{6}}\right)^2\right)^{3/2} \,.
\end{equation}
Indeed, expanding the `bottom-up' results of \cite{Dorronsoro_2017} at large scale factor, one finds that at leading order their action exactly matches this expression.
Hence the top-down origin of the wave function ultimately slightly impacts its predictions, and one might expect this to be the case for the spectral properties of fluctuations too.  In \cref{fig:NBmeasure} we compare the standard no boundary measure with the holographic measure obtained here.
We see that the holographic measure is somewhat more spread out over inflationary histories. 

\section{Discussion}
\label{secdiscussion}

We have identified a one-parameter set of complex homogeneous mass deformations of ABJM that are dual to an interior bulk describing the birth of an eternally inflating universe in the no-boundary state. The bulk saddles are solutions of a consistent truncation of supergravity in four dimensions in which the six independent complex scalars behave as a single effective inflaton. The single free parameter is related to the value of the source in the dual and governs the amount of scalar field inflation in the bulk, before the inflaton rolls down to the minimum of a positive effective cosh-potential. It would be interesting to study the eleven-dimensional uplift of these inflationary saddles, and to interpret some of the particularities of the four-dimensional solutions from an 11D viewpoint.

The saddles we have considered are so-called fuzzy instanton solutions of Euclidean supergravity, in which the scalars have a complex profile. Given we have identified a domain in which these solutions behave like Lorentzian cosmologies, it would be interesting to better understand whether there is a consistent Lorentzian theory of quantum gravity with these  cosmologies as saddle-point solutions, and whether this theory is something like an exotic star supergravity\cite{Hull:1998vg,Liu:2003qaa,Dijkgraaf:2016lym} as previously suggested in \cite{Hertog:2017ymy}.

The dependence of the partition function on the value of the external source specifies a holographic no-boundary measure on this minisuperspace of homogeneous and isotropic, inflationary, asymptotically de Sitter universes.
This holographic no-boundary measure differs slightly from the standard no-boundary measure in a cosh potential because of the presence of an additional surface term required by supersymmetry. Presumably this term also affects the spectral properties of fluctuations, showing once again that the microscopic origin of inflationary theory may carry phenomenological implications indeed.

Perhaps the most surprising property of the holographic no-boundary wave function in this model is that it contains, at least with the fermionic sector integrated out, a subleading contribution with the characteristics of the tunneling wave function. This feature can be traced to the fact that the dual adopts so-called mixed boundary conditions. The net result is that the usual no-boundary saddles provide the dominant contribution to the wave function, albeit with a slightly altered weighting due to the SUSY surface term, but that there is a subleading ’tunneling’ contribution.

The requirement that the wave function predicts classical cosmological evolution in the large volume regime constrains the range of the inflaton to values relatively close to the minimum of the potential. In fact, the potential becomes very steep for large field values, violating the conditions for inflation. In the large-field regime there are no proper no-boundary saddles that ensure the configuration on the final boundary is dynamically cobordant to nothing. Hence also swampland principles, specifically the dynamical cobordism conjecture, suggest that this regime should be excluded from the configuration space. Interestingly, we have seen that precisely the same constraint on the configuration space of the wave function follows from holographic considerations, viz. the behavior of the dual partition function. Furthermore, the constrained range of field values is roughly the Planck scale, very much in line with the swampland distance conjecture \cite{Ooguri:2006in,Agrawal:2018own}. Finally, also the fluctuation convergence criterion recently revived by Kontsevich, Segal and Witten fails for the saddles in most of the large-field regime.

Thus we arrive at an intriguing web of connections between swampland considerations such as the dynamical cobordism and distance conjectures, the conditions for the wave function to predict classical behavior in geometry and fields, EAdS/CFT, and the fluctuation convergence criterion. It is tempting to speculate that this quadrality may hold beyond the specific set-up we have considered, and that a better understanding of how these fit together may offer important insights in the fundamentals of cosmological theory.

\section*{Acknowledgements}

We thank Oliver Janssen for early collaboration and discussions, and Nikolay Bobev, Joel Karlsson, Klaas Parmentier, and Thomas Van Riet for valuable discussions. TH acknowledges support from the Flemish inter-university
project IBOF/21/084 and the PRODEX grant LISA - BEL (PEA 4000131558). JP acknowledges support from NCCR-SwissMAP. VV is supported by funding from an STFC consolidated grant, grant reference ST/X000761/1.

\bibliographystyle{utphys}

\bibliography{refs}

\end{document}